\documentclass[a4paper, amsfonts, amssymb, amsmath, reprint, showkeys, nofootinbib, twoside]{revtex4-1}
\usepackage[english]{babel}
\usepackage[utf8]{inputenc}
\usepackage[colorinlistoftodos, color=green!40, prependcaption]{todonotes}
\usepackage{amsthm}
\usepackage{mathtools}
\usepackage{physics}
\usepackage{xcolor}
\usepackage{graphicx}
\usepackage[left=23mm,right=13mm,top=35mm,columnsep=15pt]{geometry} 
\usepackage{adjustbox}
\usepackage{placeins}
\usepackage[T1]{fontenc}
\usepackage{lipsum}
\usepackage{csquotes}

\usepackage[pdftex, pdftitle={Article}, pdfauthor={Author}]{hyperref} % For hyperlinks in the PDF
\usepackage{orcidlink}
\begin{document}

\title{Comment on "Non-linear interactions in cosmologies with energy exchange"}

\author{Naman Soni\protect\orcidlink{0009-0006-7266-7893}}
    \email[Correspondence email address: ]{naman22@iiserb.ac.in}% Your name
    \affiliation{Department of Physics, Indian Institute of Science Education and Research Bhopal,
Bhopal - 462066, India}

%\date{\today} % Leave empty to omit a date

\begin{abstract}
This work provides a critical reassessment of the cosmological models presented in the article "Non-linear interactions in cosmologies with energy exchange" Eur. Phys. J. C 80, 120 (2020). It points out and corrects several mathematical inconsistencies in the original article. These include a flawed simplification of a Liénard-type equation, unjustified omissions of integration constants, and an incorrect use of the variation-of-parameters method. By deriving the exact analytical solutions, it is shown that the corrected mathematical framework fundamentally contradicts the claims of the original article.
\end{abstract}


\maketitle
\thispagestyle{plain}
\pagestyle{plain}

This comment points out important mathematical errors in \cite{RefA}. By re-evaluating the authors' derivations, specifically their Liénard-type equation, missing integration constants, and incorrect use of the variation of parameters, we show that the algebraic derivations yield different results. We show that when these equations are corrected, the main physical claims of the original work turn out to be inaccurate.

The scheme of the corrections is as follows:\\
\textbf{1.} Following the steps done in section 3 of \cite{RefA} the function \textit{g(u)} obtained in equation (14) of \cite{RefA} should be 
\begin{equation}
    g(u)=kA^2u+\frac{3}{2}kABu^2+\frac{1}{2}kB^2u^3 ,
\end{equation}
which follows the exactly integrable Liénard equation \cite{RefB} i.e. equation (15) of \cite{RefA} to be 
\begin{equation}
   u'' + (A + Bu)u' + \left(kA^2u + \frac{3}{2} kA B u^2 + \frac{1}{2} k B^2 u^3 \right) = 0 ,
\end{equation}
this form (2) is also consistent with equations (9) and (13) of \cite{RefA}. However, this does not affect the later steps. 

\textbf{2.} Moving to equations (22) and (23) of \cite{RefA}, the authors claim to have found the asymptotic solution at early times to be
\begin{equation}
    u^2 = \frac{2}{k} (\exp[B\tilde{C}^{-1}(x - x_0)] - A^2) ,
\end{equation}
and time evolution of scale factor to be
\begin{equation}
    a(t) \sim \exp \left( t^{B\tilde{C}^{-1}/2} \right).
\end{equation}
However, this solution is inconsistent with the used form of $f(u)=A+Bu$ and the appropriate solution comes out to be
\begin{equation}
    u^2=\frac{2A\tilde{C}^{-1}}{B}(x-x_1) ,
\end{equation}
where $\quad x_1=x_0-\frac{A}{B\tilde{C}^{-1}}ln(A)$.
Followed by
\begin{equation}
     a(t) \sim \exp \left[(lnt)^\frac{3}{2}\right] ,
\end{equation}
making the associated claims invalid.

\textbf{3.} The parametric time-evolution equation (25) obtained using equation (17) of \cite{RefA} in the limit of large $w$ does not mathematically follow from the preceding equations, which alters the subsequent steps in that subsection.
The proper form of the asymptotic solution is the following quadratic equation
\begin{equation}
    Au+Bu^2=\frac{\tilde{C}^{-1}}{k} ,
\end{equation}
Let the solution of (7) be $u_{\infty}$, the scale factor turns out to be
\begin{equation}
    a(t) \sim t^{u_{\infty}} .
\end{equation}

\textbf{4.} As we go through subsection 3.2 of \cite{RefA}, the assumption that the constant term $\frac{1}{D}$ in equation (35) is negligible is invalid for the case of $a\to \infty$ as this constant term will dominate since the other two terms are raised to the expressions which are essentially negative. The correct solution for  $a\to \infty$ is:

\begin{equation}
    u^2\to \frac{1}{D} ,
\end{equation}
\begin{equation}
    a\propto t^{\frac{1}{\sqrt{D}}} .
\end{equation}

\textbf{5.} In the curved case scenario i.e. section 4 of \cite{RefA}, the form of particular solution $b_p(\eta)$ seems to have been obtained using the method of variation of parameters \cite{RefC}, however, this solution is incorrect, if we follow this method we get
\begin{equation}
\begin{split}
b_p(\eta) &= C \sin(\omega_0 \eta) \int m(\eta) \cos(\omega_0 \eta) d\eta \\
&\quad - C \cos(\omega_0 \eta) \int m(\eta) \sin(\omega_0 \eta) d\eta ,
\end{split}
\end{equation}
where
\begin{equation}
C = \frac{(2-3\Gamma)(\Gamma-\gamma)}{\sqrt{2k(2-3\gamma)(2-3\Gamma)}} ,
\end{equation}
and 
\begin{equation}
\omega_0 = \sqrt{\frac{k}{2}(2-3\gamma)(2-3\Gamma)} \quad ,
\end{equation}
Regardless of this, the further work uses a slightly different method i.e. method of undetermined coefficients \cite{RefC}, here also the obtained solution i.e. equation (49) of \cite{RefA} has a descrepancy and the correct form should be

\begin{equation}
\begin{split}
b(\eta) &= c_1 \cos \left[ \sqrt{(2-3\Gamma)(2-3\gamma)k/2}(\eta - \eta_0) \right] \\
&\quad + \frac{(\gamma - \Gamma)m_0 e^{-\beta\eta}}{k(3\gamma - 2) - \left(\frac{2}{2-3\Gamma}\right)\beta^2} .
\end{split}
\end{equation}

In conclusion, our re-evaluation identifies several mathematical discrepancies in the original work which undermine the major physical statements. When the differential equations are solved in the right way, the solutions are not the ones given. Because these basic equations are broken at multiple stages, the the paper's overall conclusions about this cosmological model do not follow from the corrected equations.

\section*{Statements and Declarations}

\begin{small}
\noindent \textbf{Acknowledgements:} The author would like to thank Sukanta Panda for recommending working on the concerned article and Rajat Kumar Panda for his valuable discussions with author while carrying out this work.

\vspace{0.5cm}
\noindent \textbf{Data Availability Statement:} This is a theoretical work and this manuscript has no associated data.

\vspace{0.5cm} 
\noindent \textbf{Code Availability Statement:} This manuscript has no associated code.

\vspace{0.5cm} 
\noindent \textbf{Competing Interests:}
The author has no relevant financial or non-financial interests to disclose.

\vspace{0.5cm} 
\noindent \textbf{Funding:}
The author did not receive funding from any organization for the submitted work.

\end{small}

%\appendix*
%\input{sections/appendix1.tex}

\end{document}